\documentclass{tMAM2e}

\usepackage{epstopdf}
\usepackage[bookmarksnumbered,colorlinks,bookmarks,citecolor=blue,linkcolor=blue,urlcolor=blue,breaklinks,linktocpage]{hyperref}
\usepackage[all]{hypcap}
\usepackage{subfigure}
\usepackage{color}
\usepackage{tgpagella}
\usepackage{helvet}

\begin{document}

\makeatletter
\renewcommand{\ps@title}{
  \let\@oddhead\@empty
  \let\@evenhead\@empty
  \let\@oddfoot\@empty
  \let\@evenfoot\@empty
}
\def\draftnote{}
\makeatother

\title{Rhythmic segment analysis:\\
Conceptualizing, visualizing, and measuring rhythmic data}

\author{
\name{Bas Cornelissen\textsuperscript{a}$^{\ast}$\thanks{$^\ast$Email: mail@bascornelissen.nl}}
\affil{\textsuperscript{a}Institute for Logic, Language and Computation,\\
University of Amsterdam, Amsterdam, The~Netherlands}
\received{\today}
}

\maketitle

\begin{abstract}
This paper develops a framework for conceptualizing, visualizing, and measuring regularities in rhythmic data.
I propose to think about rhythmic data in terms of interval segments: fixed-length groups of consecutive intervals, which can be decomposed into a duration and a pattern (the ratios between the intervals).
This simple conceptual framework unifies three rhythmic visualization methods and yields a fourth: the pattern-duration plot.
When paired with a cluster transition network, it intuitively reveals regularities in both synthetic and real-world rhythmic data. 
Moreover, the framework generalizes two common measures of rhythmic structure: rhythm ratios and the normalized pairwise variability index (nPVI). 
In particular, nPVI can be reconstructed as the average distance from isochrony, and I propose a more general measure of anisochrony to replace it. 
Finally, the novel concept of quantality may shed light on wider debates regarding small-integer-ratio rhythms.
\end{abstract}


\noindent
Without rhythm, no music. 
Yet it appears that the field lacks a compelling conceptual framework for reasoning about rhythmic data. 
Suppose we are presented with a sequence of time intervals between drum-strokes. 
How might we determine whether regularities are present? How might we characterize them? 
Even the ``most empirically suitable approach to quantify rhythmic structure'' \citep{Ravignani2017MusicPercept}, the normalized pairwise variability index (nPVI), has received substantial methodological criticism \citep{Condit-Schultz2019MusicPercept} and is rather limited in what structure it can describe: only isochrony, as we will see. 
A series of recent studies, some at the interface with bioacoustics, has been addressing this gap by proposing novel visualizations and measures of rhythmic structure \citep{Burchardt2025OSF, Jadoul2025Vibration, Ravignani2017MusicPercept, Roeske2020CurrBiol}. 

The starting point for this paper is a simple observation. 
All of these studies analyze the same objects: interval pairs. 
That means that multiple visualization methods (phase, ratio, and raster plots) and rhythmic measures (rhythm ratios and the nPVI) can be unified in a single theoretical framework that generalizes the notion of an interval pair. 
The goal of this paper is to develop that framework, and nothing more. 
Since the focus is narrow, I will not review the literature at large but refer to other sources for broader surveys \citep{Burchardt2025OSF,Jadoul2025Vibration,VanHandel2022}. 

This paper, in short, addresses how to (1) conceptualize, (2) visualize, and (3) measure regularities in rhythmic data. 
The approach I put forward, \emph{rhythmic segment analysis}, is domain agnostic and does not rely on music-specific assumptions like a meter.
Instead, it conceptualizes rhythmic data solely in terms of local rhythmic relationships, by breaking it down into fixed-length interval sequences. 
These \emph{segments} can be decomposed into a \emph{duration} and a \emph{pattern}: the ratios between the intervals. 
Several existing visualization methods show either segments or patterns, and I propose a novel method, the pattern-duration plot, to visualize both.
When overlaid with a network summarizing segment transitions, it can intuitively reveal rhythmic regularities in a dataset. 
Finally, this framework naturally incorporates two measures of rhythmic regularity. 
First, rhythm ratios correspond directly to length-2 patterns. 
Second, the nPVI naturally emerges as an average distance from isochrony, and leads to a more general (and normalized) measure of anisochrony that can replace the nPVI.

The strength of the proposed framework is its simplicity and generality. 
Rhythmic segment analysis naturally ties together existing work, can help clarify our reasoning about rhythmic data, and is straightforward to implement: (1) split intervals into fixed-length segments, (2) sum them, and then (3) normalize the segments to get durations and patterns. 
A small Python package, \emph{rhythmic-segments}, which I release alongside this paper simplifies things even further, especially when it comes to handling metadata. 
Installation instructions and extensive documentation are available online, as are all the data and code discussed in this paper.\footnote{
  The package \emph{rhythmic-segments} can be installed using pip, and the documentation can be found at \url{https://bacor.github.io/rhythmic-segments}.
  All data and code used in this paper, along with the supplementary materials, have been deposited at \url{https://doi.org/10.5281/zenodo.17753529}.
}

\begin{figure}
  \begin{center}
  \includegraphics[width=\textwidth]{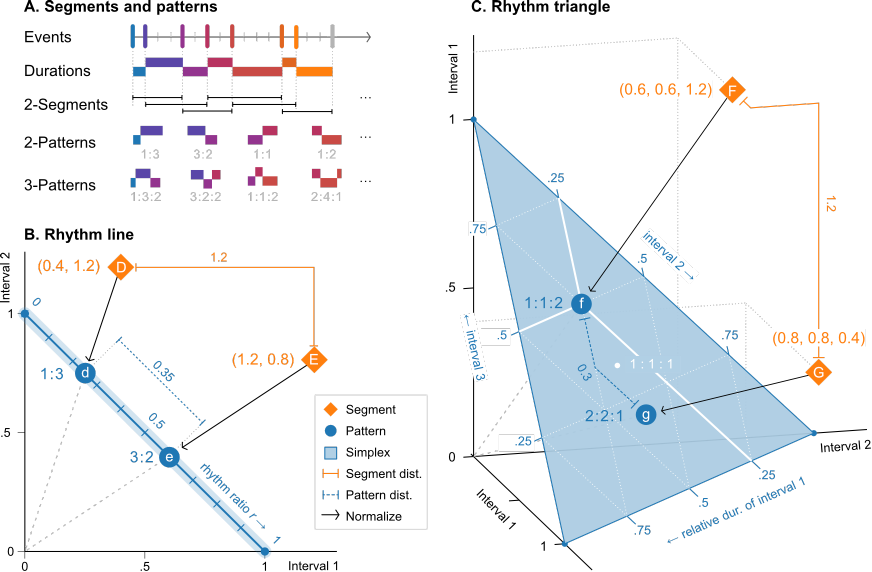}
  \caption{
    \textbf{Rhythmic segment analysis} studies fixed-length segments of a sequence of durations: interval pairs or  triplets for example. 
    Every segment has a rhythmic pattern: the ratios between its intervals (\textbf{A}).
    Length-$n$ patterns are points in a $(n-1)$-dimensional simplex.
    Panel \textbf{B} shows the space of length-2 patterns: the rhythm line.
    Values along this line are known as \emph{rhythm ratios} $r$ with isochrony corresponding to $r=0.5$.
    Normalization maps segments to their patterns, as illustrated for segments $D$ and $E$ with patterns $d$ and $e$. 
    Panel \textbf{C} shows the space of length-3 patterns, the \emph{rhythm triangle}.
    The white crosshairs shows how to read the coordinates of patterns inside the triangle: pattern $f$ has relative durations $(.25, .25, .5)$.
    Both panels show how the \emph{segment distance} is measured along the gridlines of the ambient space, while the \emph{pattern distance} is measured along the gridlines of the simplex.
  }
  \label{fig:rsa}
  \end{center}
\end{figure}

\section{Conceptualization}
\label{sec:conceptualization}

I would like to start by laying out the basic ingredients of a rhythmic segment analysis. 
To address a wide audience, I will introduce the key concepts with examples and geometric intuitions wherever possible. 

\subsection{Segments, patterns, and durations}
The starting point is ‘rhythmic data,’ by which I mean a sequence of
\emph{durations}.
In the most typical scenario, the durations are the time intervals between events such as note onsets (i.e.~inter-onset intervals).
I will therefore present the framework for a sequence of \emph{intervals}, even though it works for any sequence of durations (e.g. note durations while ignoring rests).
The motivating use case is one where durations are moreover measured in continuous time—derived from a recording rather than a score.

The idea is to analyze rhythmic data in terms of its \emph{segments}: groups of consecutive intervals obtained by sliding a fixed-length window over the intervals—interval $n$-grams, in short.
For example, the length-3 segments of the intervals $(.1, .1, .2, .2, .2, .4, .3, \dots)$ would be $(.1, .1, .2)$, $(.1, .2, .2)$, $(.2, .2, .2)$, $(.2, .2, .4)$, and so on. 
Segments like $(.1, .1, .2)$ and $(.2, .2, .4)$ have different \emph{duration}—0.4s and 0.8s respectively—but both represent the same \emph{pattern}: short, short, twice as long. 
This pattern can be described in at least two ways: using the \emph{ratios} between the intervals, $1 : 1 : 2$, or using the \emph{relative durations} $(.25, .25, .5)$ of the intervals. 
Both descriptions are equivalent and can be used interchangeably. 

\subsection{The rhythm simplex}
Patterns are effectively normalized segments.
Dividing the entries of the segment $(.1, .1, .2)$ by its duration $.4$ results in the pattern $(.25, .25, .5)$.
Because it sums to one, the last entry is fully determined by the others: $.5 = 1-.25-.25$.
Differently put, a pattern of length three has only two degrees of freedom. 
And by the same argument, patterns of length $n$ must form an $(n - 1)$-dimensional space \citep{Desain2003Perception}. 
That space of patterns is known as a \emph{rhythm simplex} \citep{Jacoby2017CurrBiol}. 
I will use the more specific terms \emph{rhythm line} and \emph{rhythm triangle} for the spaces of length-2 and length-3 patterns respectively.

Figure~\ref{fig:rsa}B shows the rhythm line as a diagonal line segment.
The position along that line, expressed as a number between 0 and 1, is known as the \emph{rhythm ratio} $r$ or $r_k$ \citep{Roeske2020CurrBiol}. 
It is the relative duration of the first interval in a length-2 segment, $r(a, b) = a / (a+b)$.
A point on the rhythm line, such as point $e$ in Figure~\ref{fig:rsa}B, can thus be described in three equivalent ways: as the ratio $3:2$, as the relative durations $(0.6, 0.4)$, or as the first number in that pair: the rhythm ratio $r=0.6$. 
Of particular importance is the value $r = 0.5$, which indicates the isochronous pattern $(0.5, 0.5)$ or $1 : 1$,  where all intervals have the same duration. 
Figure~\ref{fig:rsa}B shows how normalization maps a segment onto the simplex. 
As a result, all segments on a single line through the origin represent the same pattern. 
The farther they lie from the origin, the slower their tempo. 

Except for the rhythm ratio,\footnote{
  One could of course generalize the rhythm ratio to higher dimensions as the first $n-1$ relative durations, $r(p_1, \dots, p_n) = (p_1, \dots, p_{n-1})$, but it would lose its attraction as a scalar quantity—and we do not need it here.
} all this generalizes to higher dimensions, as Figure~\ref{fig:rsa}C shows for the rhythm triangle.
This triangle was first introduced as the \emph{chronotopic map} by \citet{Desain2003Perception} in a study of categorical rhythm perception. 
More recently, it featured prominently in studies of perceptual priors of listeners from different cultures \citep{Jacoby2017CurrBiol,Jacoby2024NatHumBehav}.
In the center of the simplex, we find the isochronous pattern $(\tfrac{1}{3}, \tfrac{1}{3}, \tfrac{1}{3})$ or $1 : 1 : 1$. 
As one approaches the corners, the patterns become increasingly non-isochronous, approaching impossible patterns like $1 : 0 : 0$ where one interval fills the entire segment and the two others are vanishingly small.

\subsection{Measuring distances}
For length-4 patterns, the rhythm simplex is a tetrahedron. 
Beyond that, the space becomes increasingly hard to visualize. 
But longer segments and patterns can also be studied using suitable distance metrics. 
Instead of Euclidean distance, the \emph{segment distance} I propose is the sum of the absolute differences between two segments $x = (x_1, \dots, x_n)$ and $y = (y_1, \dots, y_n)$,
\begin{align}
  d_s(x,y) = |x_1 - y_1| + \cdots + |x_n - y_n|.
\end{align}
This measures the total adjustment needed to turn one segment into the other. 
For example, the segment distance between $(1, 2, 3)$ and $(4, 5, 6)$ is $|1 - 4| + |2 - 5| + |3 - 6| = 9$—seconds, perhaps. 
Geometrically, the segment distance is the length of the shortest path between two segments, provided the path follows the grid lines (see Figure~\ref{fig:rsa}). 
For that reason, it is also known as the \emph{taxicab distance}: it measures the path a taxi would take in a grid-like city. 
(Note that the norm of a segment under the corresponding $L_1$-norm, is precisely its duration $\|x\|_1 = x_1 + \dots + x_n$.)

One could compare patterns in the same way.
However, because patterns are normalized, the segment distance would count all differences \emph{twice}.
For example, the patterns $(0.25, 0.75)$ and $(0.75, 0.25)$ are $0.5 + 0.5 = 1$ apart according to the segment distance. 
But when increasing the first interval by $0.5$, normalization automatically implies an equal change to the second interval.
Therefore, using \emph{half} of the segment distance seems more appropriate.
Formally this measure is known as the \emph{total variation distance}, but I will call it the \emph{pattern distance}:
\begin{align}
  d_p(p, q) = \tfrac{1}{2} \big( |p_1 - q_1| + \cdots + |p_n - q_n| \big)
\end{align}
The pattern distance also measures the shortest path between two patterns, but now it must follow the grid of the simplex (Figure~\ref{fig:rsa}). 
Moreover, it is itself normalized, so that the distance between any two corners of the simplex is 1.

\begin{figure}
  \begin{center}
  \includegraphics{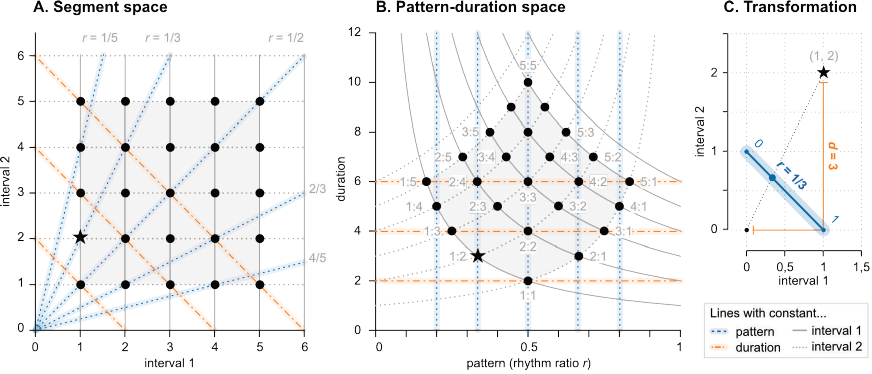}\\
  \caption{
    \textbf{Pattern-duration space.}
    Segment can be represented either in segment space (\textbf{A}) or in pattern-duration space (\textbf{B}), here by plotting the rhythm ratio horizontally and the duration vertically.
    This explicitly isolates tempo (vertically) from the rhythmic content (horizontally). Moving between the two descriptions is a coordinate transform similar to polar coordinates (\textbf{C}): the pattern $r$ determines the direction, the duration $d$ the distance from the origin.
    A square of segments (black dots) is plotted in both spaces to illustrate the transformation, along with lines of constant pattern, duration, and interval. 
    For example, all points on a dashed blue line represent the same pattern, and lines through the origin in phase space become vertical lines in pattern-duration space.
    Labels in (\textbf{B}) show the coordinates of the segments in segment space.
  }
  \label{fig:coordinate-transform}
  \end{center}
\end{figure}

\subsection{Pattern-duration space}

We started this section by noting that a segment $(x_1, \dots, x_n)$ can be decomposed in a pattern $p=(x_1/d, \; \dots, \; x_n/d)$ and a duration $d=x_1+\dots+x_n$.
Obvious as it may be, this has conceptual significance.
The pattern-duration decomposition effectively separates tempo from the rhythmic ‘content’.  
The rhythmic content (i.e.~the pattern) describes how intervals in a segment are related irrespective of their absolute durations—similar perhaps to how relative pitch might describe relations between notes irrespective of their absolute frequencies. 
And so it may well be useful to conceive segments not as sequences of intervals, but as a pattern-duration pairs $(p, d)$.

For segments of length 2 the resulting \emph{pattern-duration space} can easily be visualized by plotting duration against the rhythm ratio, as illustrated in Figure~\ref{fig:coordinate-transform}B.
Normalization is essentially the transformation that maps segments to pattern-duration pairs, and multiplication recovers the original segment $x = p \cdot d$. 
Intuitively, the transform rotates the segment space by 45 degrees and blows up the area close to the origin. 
But it has a more precise geometric interpretation.
Recall that all segments with the same pattern fall on a line through the origin.
A pattern effectively describes the direction of that line, while the duration specifies the distance from the origin along that line (see Figure~\ref{fig:coordinate-transform}C).
The pattern and the duration behave like the angle and radius in polar coordinates—exactly so, in fact: in taxicab geometry, the simplex \emph{is} the unit ball on the positive part of space.
Moving to pattern-duration space, differently put, is a polar coordinate transform.

In summary, we have introduced three related spaces.
First, the \emph{segment space} of interval $n$-tuples with the segment distance.
Second, the \emph{pattern space}, the simplex of normalized $n$-tuples with the pattern distance.
Third, the \emph{pattern-duration space} is a coordinate transform of the \emph{segment space} that separates tempo from rhythmical content.

\begin{figure}
  \begin{center}
  \includegraphics{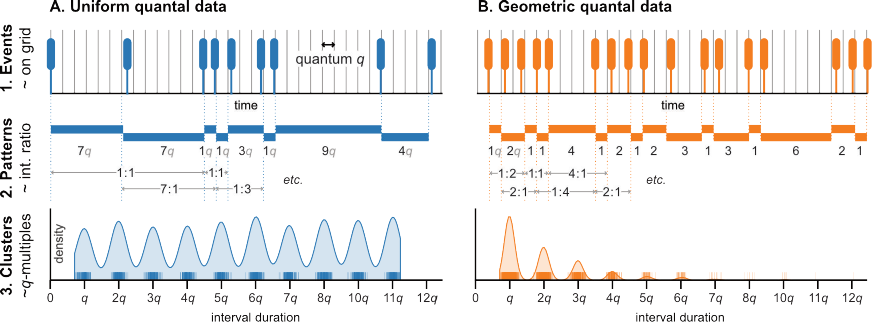}
  \caption{
    \textbf{Quantality.}
    A rhythmic dataset is quantal when intervals tend to be integer multiples of a fixed quantum $q$ (row 3).
    Intervals between events that approximately fall on a grid (row 1) are an example of quantal data.
    Intervals in a quantal dataset are approximately discrete, and since they are all multiples of a quantum, must be related by integer ratios (row 2).
    These three phenomena are illustrated for two types of random data: noisy multiples of a quantum, with the multiples drawn from either (\textbf{A}) a uniform distribution or (\textbf{B}) a geometric distribution, which favour short-intervals.
  }
  \label{fig:quantality}
  \end{center}
\end{figure}

\subsection{Quantality}

To end this section, let me propose the concept of \emph{quantality} to describe the approximate discreteness typical of musical rhythm.
I will call rhythmic data \emph{quantal} when the interval distribution consists of clusters close to integer multiples of a smallest temporal unit, the \emph{quantum}.
That means that intervals are of the form $m\cdot q + \epsilon$ for an integer $m$, a fixed quantum $q$ and some noise $\epsilon \ll q$: they are noisy renditions of quantum multiples.
The third row of Figure~\ref{fig:quantality} illustrates this with random intervals, drawing the multiples $m$ either uniformly from $\{1, 2, \dots, 11\}$, or from a geometric distribution to favour short multiples.
I deliberately opted for the neutral, domain-general term \emph{quantum}, to avoid the theoretical baggage of \emph{beat}, \emph{subdivision}, \emph{tatum}, and so on.
Moreover, depending on the data, the quantum could be any of those. 
Finally, I avoid the term \emph{quantization} (discretizing continuous intervals to a grid) because the notions are not equivalent: quantal data are generally \emph{not} quantized, but quantized data are always quantal. 

Fully developing the concept of quantality is beyond the scope of this paper, and for that reason the definition is relatively informal. 
However, I do want to relate it to two other phenomena.
First, if a sequence of events is \emph{grid-like}, meaning that all events roughly fall on a grid (Figure~\ref{fig:quantality}, top row), then all intervals are multiples of the grid size, and so the intervals must be quantal.
Second, ratios between intervals in a quantal dataset will tend to be \emph{integer ratios}: the ratio between any two approximate quantum multiples $mq + \epsilon$ and $kq + \delta$ is close to an integer ratio, namely $m:k$.\footnote{
  Note that neither of the converse implications is strictly speaking true.
  First, quantal data need not be grid-like: the error term can push events off the grid, even if the interval durations approximate quantum multiples.
  Second, consider the intervals $1, 1, 1, \dots, 1, \sqrt{2}, \sqrt{2}, \dots, \sqrt{2}$.
  The relations between successive intervals is always an integer ratio (namely, $1:1$, with one exception), but there is no quantum $q$ of which $1$ and $\sqrt{2}$ are integer multiples.
  However, one can define a notion of \emph{integer-ratio segment clustering} that \emph{is} equivalent to quantality: when segment clusters tend to have integer ratio patterns \emph{and} their cluster durations are integer multiples of a quantum. In this case, segment clusters essentially form a grid.
}
Musically, this should make sense: if a rhythm falls on a 16th-note grid, all intervals are roughly multiples of the 16th note.
In other words, the rhythm is quantal. 
That in turn means that all intervals are roughly related by integer ratios. 
To the extent that multiples are small, the ratios will be \emph{small integer ratios} (see Figure~\ref{fig:quantality}, A2 vs.~B2).

\section{Visualization}
\label{sec:visualization}

Rhythmic segments and patterns can have arbitrary length, but only pairs and triplets can be effectively visualized in two dimensions without resorting to dimensionality reduction techniques. 
It turns out that three existing plotting methods can be naturally understood as segment or pattern plots, to which the previous section added a fourth: the pattern-duration plot.
We now compare these different visualization methods using three synthetic datasets, since we can fully understand and control their properties, before turning to real-world data.

The first dataset samples intervals uniformly between $0.2$ and $2$s.
The second is the geometric quantal data from Figure~\ref{fig:quantality}B: integers from a geometric distribution, scaled by a quantum of $0.2$s, with Gaussian noise.
The third repeats a fixed template of integer intervals, adds Gaussian noise in the same way, and then scales the result by a quantum of $0.5$s.
All three datasets are continuous, but only the last two are quantal: those contain, by approximation, a limited number of possible intervals and therefore exhibit clear rhythmic categories.

\begin{figure}
  \begin{center}
  \includegraphics[width=\textwidth]{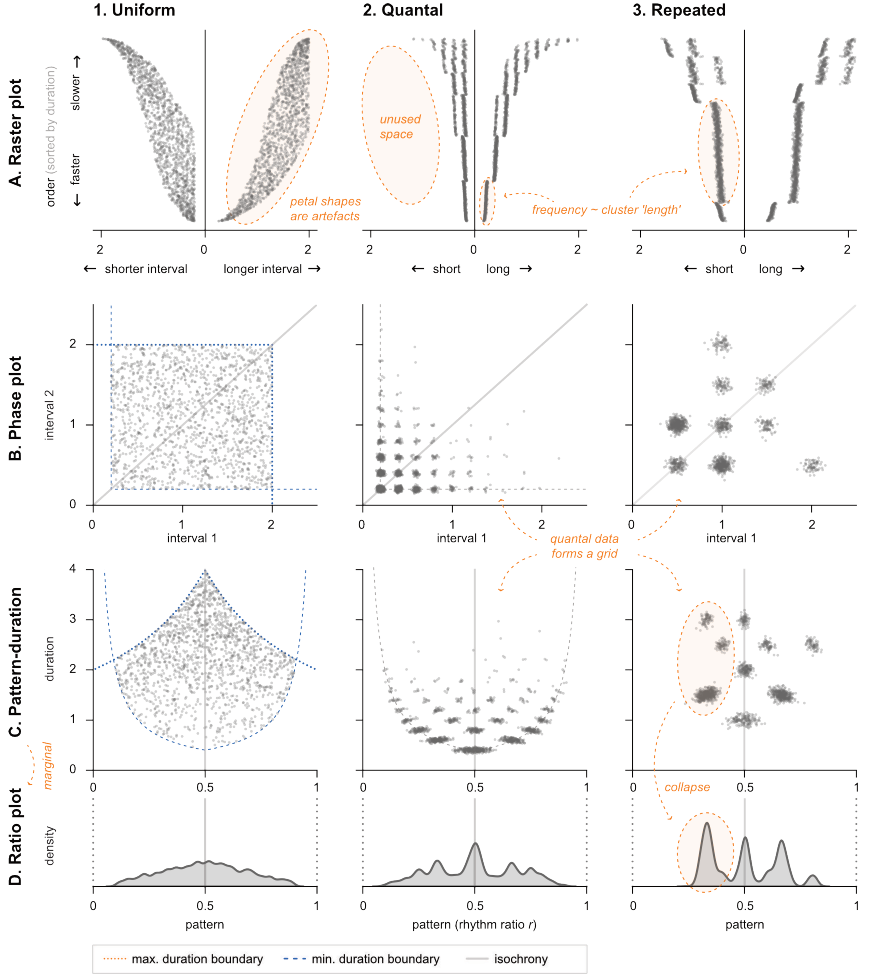}
  \caption{
    \textbf{Four visualization methods.} 
    Columns show three types of synthetic data: (1) uniform intervals, (2) geometric quantal intervals, and (3) a noisy, repeated rhythm. 
    (\textbf{A}) Raster plots  sort segments vertically by their duration; the shorter interval of each segment is plotted to the left, the longer one to the right. 
    (\textbf{B}) Phase plots  plot the first against the second interval. 
    (\textbf{C}) Pattern-duration plots show the rhythm ratio horizontally and the duration. 
    (\textbf{D}) A ratio plot is a marginal density plot of a pattern-duration plot.
    See the main text for details.
  }
  \label{fig:plots}
  \end{center}
\end{figure}

\subsection{Raster, phase, and ratio plots}
\emph{Raster plots}, shown in Figure~\ref{fig:plots}, are the most intriguing and elusive of these plots \citep{Roeske2020CurrBiol}. 
Segments are sorted vertically by their duration, so that the slowest segments are on top and the fastest at the bottom. 
Each segment is then plotted using two points: the shorter interval on the left and the longer interval on the right. 
Raster plots produce beautiful flower-like figures with richly patterned petals. 
The flower shape, however, is an artefact. 
It appears even for uniform data, demonstrating that raster plots make inefficient use of visual space. 
Moreover, interpreting the patterning inside the petals is difficult, hampered by the fact that the vertical axis has no scale and that sorting blows up frequent segments, resulting in very long clusters.

\emph{Phase plots} or \emph{lag plots}, shown in Figure~\ref{fig:plots}B, present a more interpretable alternative, as we have in fact already seen in the previous section \citep{Ravignani2017MusicPercept,Ravignani2016NatHumBehav}.
The first interval of each segment is shown along the horizontal axis, the second interval along the vertical axis. 
When the data contain recurrent regularities, connecting successive segments with lines will produce distinct geometric figures \citep{Ravignani2017MusicPercept}.
I have omitted those lines in Figure~\ref{fig:plots}B, as they would also obscure the segment clusters, but you will later see them in the background of Figure~\ref{fig:cluster-transition-network}.
Finally, Figure~\ref{fig:plots}B-2 shows that segments clusters of quantal data tend to form a grid.

\emph{Ratio plots}, shown in Figure~\ref{fig:plots}D, visualize the distribution of rhythm ratios over the rhythm line using a kernel density estimate \citep{Roeske2020CurrBiol}. 
Ratio plots have been used in studies of categoricity in rhythmic data \citep[e.g.][]{DeGregorio2021CurrBiol}, but they can misrepresent the clustering structure of the data. 
For example, the ratio plot suggests that the repeated dataset has around three rhythmic categories (Figure~\ref{fig:plots}D-3), while the phase plot shows that there are many more segment clusters (Figure~\ref{fig:plots}B-3). 
The underlying reason becomes evident when looking at pattern-duration plots.

\emph{Pattern-duration plots}, shown in Figure~\ref{fig:plots}C, show the duration of each segment vertically, against its pattern (the rhythm ratio) horizontally, as also explained in the previous section.
This means that a ratio plot shows the marginal density of a pattern-duration plot.
As a result, a ratio plot can collapse multiple segment clusters into a single mode, even when the segment clusters have distinct patterns. 
For example, on the left side of Figure~\ref{fig:plots}C-3 
we see clusters with patterns  $1 : 2$ and $2 : 3$ that are collapsed into a single mode in Figure~\ref{fig:plots}D-3.

\subsection{Pattern-duration plots}

Pattern-duration plots are closely related to the \emph{tempo-ratio plots} in \citet{Roeske2020CurrBiol}, which show only half of the rhythm line. 
However, pattern-duration plots need not be symmetric around isochrony ($r = 0.5$). 
This can be seen from Figure~\ref{fig:plots}C-3, but an additional example may be informative. 
Looping the intervals $(1, 3, 9)$ with some noise will produce three clusters of segments around $(1, 3)$, $(3, 9)$, and $(9, 1)$. 
In a pattern-duration plot, this translates into clusters around $(0.25, 4)$, $(0.25, 12)$, and $(0.90, 10)$. 
The mirror images around $(0.75, 4)$, $(0.75, 12)$, and $(0.10, 10)$ are all absent. 
This argues against omitting half of the simplex, as in the tempo-ratio plots of \citet{Roeske2020CurrBiol}.

Pattern-duration plots may show a curved \emph{minimum-duration boundary} at the bottom (dashed line in Figure~\ref{fig:plots}C-1). 
The empty space below the boundary maps to the empty bands next to the axes in a phase plot (Figures~\ref{fig:plots}B-1 and \ref{fig:plots}C-1). 
Such a boundary appears when the dataset has a shortest interval, perhaps resulting from a production or measurement constraint. 
Any segment below this boundary must contain an interval shorter than the minimum duration, which is impossible by definition. 
Similarly, one can find a \emph{maximum-duration boundary} at the top (dotted line in Figure~\ref{fig:plots}C-1).

\begin{figure}
  \begin{center}
  \includegraphics[width=\textwidth]{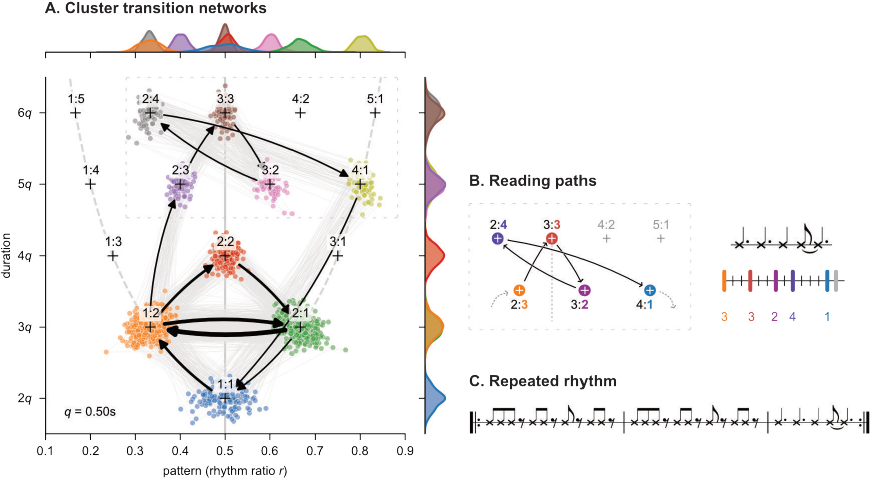}
  \\
  \caption{
    \textbf{A cluster transition network.} 
    (\textbf{A})
    Noisy repetitions of a fixed rhythm are shown as a pattern-duration plot with marginal density plots. 
    This dataset is quantal and so durations (vertical axis) have been expressed as multiples of the quantum $q=0.5$s. 
    Colors indicate clusters, and a cluster transition network shows transitions between clusters, with the thicker lines indicating more frequent transitions. 
    Gray lines in the background show individual segment transitions.
    All possible segments are marked and annotated: $2 : 3$, for example, indicates the segment $(1.0, 1.5)$ consisting of two and three quanta. 
    (\textbf{B}) Reading only the second of the annotated numbers, gives the rhythm produced by a path through the network.
    This is illustrated for the final bar of the repeated rhythm shown in (\textbf{C}).
  }
  \label{fig:cluster-transition-network}
  \end{center}
\end{figure}

Quantal data with short intervals (i.e.~small-integer multiples) tends to produce clusters along the minimum-duration boundary (see Figure~\ref{fig:plots}C-2 and Figure~\ref{fig:cluster-transition-network}).
After all, quantal data has a minimum duration by definition: the quantum. 
Its minimum-duration boundary contains segments in which one of the intervals equals one quantum (and the other some multiple of it). 
Moreover, if there is a minimum duration, patterns far from isochrony (e.g. $0.1$ or $0.9$) can necessarily only appear in segments with a relatively long duration, explaining why the scatter plots often widen toward the top. 
Finally, note that these duration boundaries arise from constraints on interval duration. 
Constraints on segment duration produce horizontal boundaries.

\subsection{Cluster transition networks}
One drawback of scatter plots is that you lose the sequential order of segments. 
\citet{Ravignani2017MusicPercept} show sequential order in his phase plots by connecting successive segments with lines. 
These trajectories can, however, quickly clutter a plot, and so I propose instead to summarize segment transitions in a \emph{cluster transition network}. 
The idea is to group segments into clusters and visualize cluster transitions as a weighted network. 
Figure~\ref{fig:cluster-transition-network} shows a cluster transition network of the repeated rhythm also shown in Figure~\ref{fig:plots}-3, with a Ravignani-style phase plot as light gray lines in the background. 
To create the network, clusters of segments are identified using the clustering algorithm HDBSCAN \citep{McInnes20172017ICDMW,Pedregosa2011JMachLearnRes}.
The resulting clusters are then used as nodes, adding edges whenever segments from two clusters succeed one another in the data. 
Edges are weighted by transition frequency and infrequent ones ($<15$) are pruned. 
The result is a network that effectively summarizes the main sequential order.

\begin{figure}
  \begin{center}
  \includegraphics[width=\textwidth]{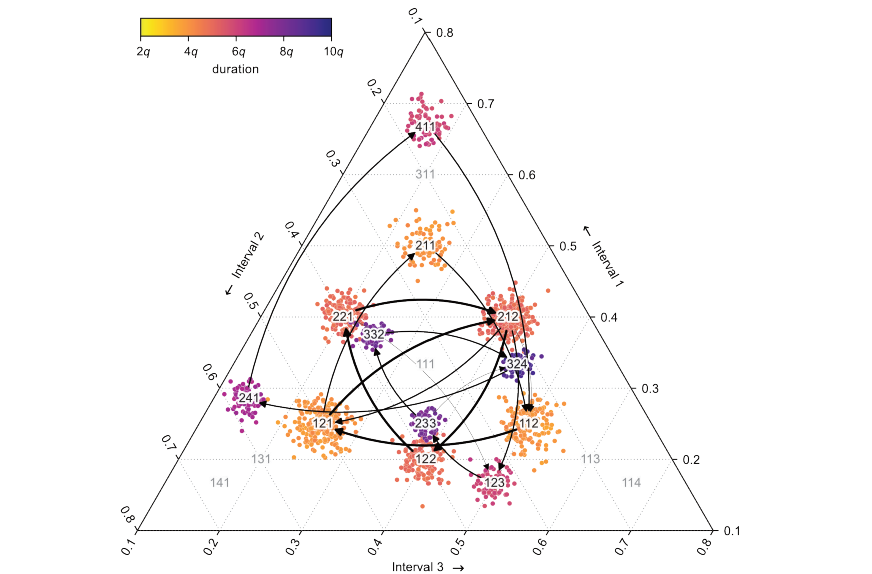}
  \\
  \caption{
    \textbf{A length-3 pattern-duration plot} of the repeated synthetic dataset. 
    The place in the rhythm triangle represents the pattern—the axes have been rotated with respect to Figure~\ref{fig:rsa}(\textbf{C})—and color encodes the segment duration, expressed in multiples of the quantum. 
    Compared to Figure~\ref{fig:cluster-transition-network}, the additional time step further disambiguates the fastest clusters.
  }
  \label{fig:3d-pat-dur}
  \end{center}
\end{figure}

Any path through this network produces a rhythm, and it is easy to determine which.
Since this is a quantal dataset, the duration axis has been rescaled to show multiples of the quantum $q = 0.5$s, and all integer-ratio segments have been annotated.
This highlights that all clusters correspond to small-integer-ratio patterns.
The path through the top section of the network, starting from $2:3$, is deterministic.
Figure~\ref{fig:cluster-transition-network}B illustrates that by reading only the second number in the annotated ratios, one can work out that the deterministic section follows a rhythm of the form $3:3:2:4:1$. 
The lower part of the network is more ambiguous and suggests multiple routes through the clusters around $1 : 1$, $1 : 2$, $2 : 2$, and $2 : 1$. 
In reality, however, the rhythm always takes the same path: it consists of two repetitions of the backbone of Steve Reich’s \emph{Clapping Music} with a coda (Figure~\ref{fig:cluster-transition-network}C). 
One way to further disambiguate the non-deterministic transitions at the bottom of the network is by repeating the analyses with longer segments.
Figure~\ref{fig:3d-pat-dur} shows a cluster transition network for length-3 segments \citep{Hunter2007ComputSciEng,Ikeda2024}.
Since the pattern space is now a two-dimensional triangle, duration is shown using colors.
With longer segments, the number of clusters will also increase. 

\subsection{Case study: Cuban salsa and son}
So far, I have only presented synthetic rhythmic data. 
To illustrate a more realistic use case of the pattern-duration plots, I turn to the IEMP Cuban salsa and son dataset \citep{Clayton2021EMR,Poole2019}.
The corpus consists of five songs in the styles of son and salsa, performed by the Cuban band Asere and contains automatically extracted onset annotations for all instruments separately. 
A length-2 pattern-duration plot was made using those onsets. 
I computed the 16th-note duration (the quantum) from the metrical cycle annotations, which allowed me to annotate integer-ratio segments as before. 
HDBSCAN is used to cluster segments, with a minimum of 10 segments per cluster. 
This clustering method can decide not to assign a segment to a cluster. 
Such segments are marked with gray plus signs and ignored when computing the network.

\begin{figure}
  \begin{center}
  \resizebox*{12cm}{!}{
    \includegraphics{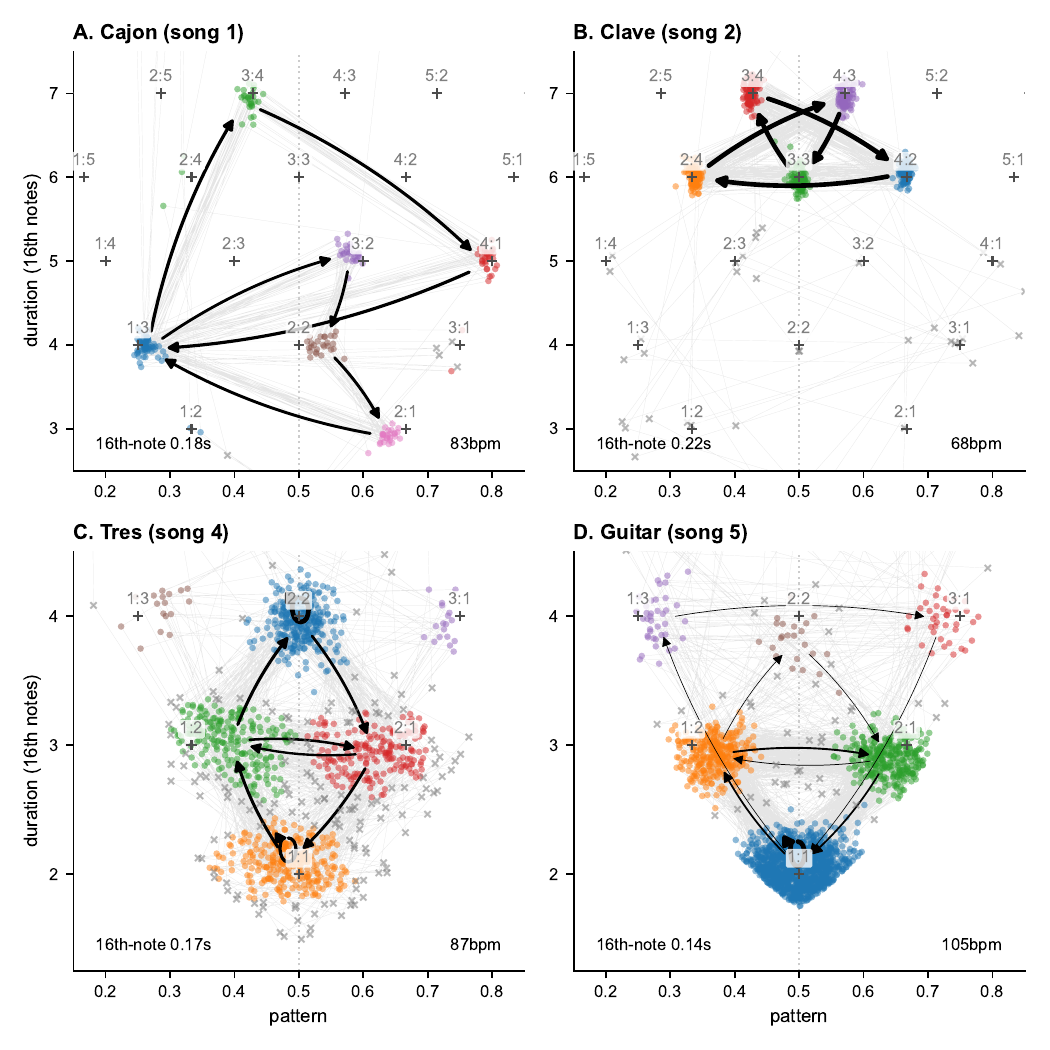}
  }\\
  \caption{
    \textbf{Selected instruments} from the IEMP Cuban salsa and son corpus. 
    Pattern-duration plots for four instruments playing either relatively slow rhythms (cajon and clave, top) or relatively fast rhythms (bottom). 
    Expressing duration (vertically) in terms of the quantum (a 16th note) makes the plots directly comparable. 
    Clusters were identified automatically as discussed in the main text. 
    Starting at $4 : 3$, the clave in panel \textbf{(B)} has the form $3:3:4:2:4$.
    }
  \label{fig:iemp-css}
  \end{center}
\end{figure}

Figure~\ref{fig:iemp-css} shows pattern-duration plots for four selected instruments across four songs.\footnote{
  Additional plots for all other instruments and songs are available in the repository alongside the code to generate them.
}
For the cajon in the first song (Figure~\ref{fig:iemp-css}A), we see two cycles representing the rhythms $1:3:4$ and $1:3:2:2$, but we cannot see how they alternate. 
Note that the second rhythm is a variant of the first (it splits the final interval in half), and observe how this transforms the path through pattern-duration space. 
The clave in song 2 (Figure~\ref{fig:iemp-css}B) performs a consistent $3:3:4:2:4$ rhythm: precisely the 3-2 son clave pattern of this song \citep{Poole2019}. 
The tres (Figure~\ref{fig:iemp-css}C) and guitar (Figure~\ref{fig:iemp-css}D) play much faster rhythms of one or two 16th notes. 
Disambiguating the precise rhythms played would again require additional analyses, for example, using longer segments.

\section{Measurement}
\label{sec:measurement}

The visualization methods introduced in the previous section are probably best paired with quantitative measures. 
In this section I demonstrate how the distance measures introduced in Section~\ref{sec:conceptualization} give rise to useful measures of rhythmic structure. 

\subsection{Anisochrony}
The most obvious quantity to measure is regularity: how isochronous is a given dataset? 
Within the proposed framework, one could measure the average distance of a set of patterns to the isochronous pattern at the center of the simplex. 
More precisely, that would measure anisochrony: how non-isochronous a pattern is. 
Formally, the \emph{anisochrony} of a pattern $(p_1, \dots, p_n)$ can be defined as the normalized distance from isochrony,
\begin{align}
\alpha(p_1, \dots, p_n) 
  = \frac{n}{2(n-1)} \sum_{i=1}^n \left|p_i - \tfrac{1}{n}\right|.
  \label{eq:anisochrony}
\end{align}
Unpacking this, the pattern distance between a pattern $p$ and the isochronous pattern $c=(\tfrac{1}{n}, \tfrac{1}{n}, \dots, \tfrac{1}{n})$ at the center of the simplex is $d_p(p, c) = \tfrac{1}{2}\sum_i |p_i-\tfrac{1}{n}|$. 
The remaining normalizing constant $n / (n - 1)$ ensures that the measure scales from $0$ in the center (perfectly isochronous) to $1$ (maximally anisochronous) in the corners of the simplex.\footnote{
  A corner is a pattern with only zeros and a single one, and so it follows that the distance from any corner to the center is $\tfrac{1}{2}[(n-1) \cdot |0 - \tfrac{1}{n}| +
    |1-\tfrac{1}{n}|]
    = \tfrac{1}{2}(\frac{n-1}{n} + \frac{n-1}{n})
    = \frac{n-1}{n}$.
} 
It may be convenient to define the anisochrony of a \emph{segment} as the anisochrony of its pattern: $\alpha(x_1, \dots, x_n) = \alpha(x_1/d, \dots, x_n/d)$, where $d$ is the duration of $x$.

For interval pairs, the above definition can be dramatically simplified. 
If two numbers $r$ and $s$ sum to one, you can show that $|r-s| = |r-\tfrac{1}{2}| + |s-\tfrac{1}{2}|$. 
Since the normalizing constant in \eqref{eq:anisochrony} disappears for $n=2$, it follows that anisochrony of a pattern $(r, s)$ is simply the absolute difference between the relative durations
\[
  \alpha(r, s) = |r-s|.
\]
For example, a pattern with rhythm ratio $r = 0.5$ has anisochrony $0$ as it is perfectly isochronous. 
A pattern with rhythm ratio $0.25$ has anisochrony $0.5$ and as the ratio approaches $0$, the anisochrony grows towards $1$.

\subsection{Reconstructing the nPVI}
The proposed measure of anisochrony turns out to completely reconstruct the normalized pairwise variability index (nPVI) \citep{Low2000LangSpeech}. 
Originally introduced to quantify speech rhythm, the nPVI became one of the standard measures of rhythmic regularity, in particular in music-language comparisons \citep[see e.g.][]{Condit-Schultz2019MusicPercept,VanHandel2022}.
For a sequence of intervals $i_1, \dots, i_K$, the nPVI is defined as 
\[
\text{nPVI} 
  = \frac{200}{K-1}  
    \sum_{k=1}^{K-1} \left| \frac{i_k - i_{k+1}}{i_k + i_{k+1}} \right|.
\]
One may recognize that the nPVI is $200$ times the average of some quantity: the term in the summation.
If we abbreviate $d=i_k + i_{k+1}$, that quantity takes the familiar form $|(i_k - i_{k+1}) / d| = |i_k/d \, - \, i_{k+1}/d|$: the absolute difference between the relative durations of the segment $(i_k, \; i_{k+1})$.
As we have shown above, this is exactly the anisochrony of segment $(i_k, i_{k+1})$ and means that the nPVI is $200$ times the average anisochrony.
In short, \emph{the nPVI measures average distance from isochrony}.
One might take this result as an argument for replacing the nPVI with the anisochrony, as the latter more conveniently scales from $0$ to $1$. 
Alternatively, one could even flip the definition to define the \emph{isochrony} as one minus the anisochrony.

I first reported this result in \citet[][Chapter 8]{Cornelissen2024}, but \citet{Jadoul2025Vibration} recently reached a closely related conclusion. 
They describe the nPVI as a summary statistic of rhythm ratios. 
To show that, it suffices to express the anisochrony of a pattern $(r, s)$ solely in terms of $r$. 
That is indeed possible: it is $2|r - 0.5|$. 
But what quantity does this represent? 
\citet{Jadoul2025Vibration} come very close to answering that question when they write that the nPVI quantifies how much the rhythm ratios ``diverge from 0.5 (i.e.~isochrony)'' on average.
It is a small step to observe that this ``divergence'' is a distance, but arguably an important one: it provides a geometric interpretation that is generalizable to longer patterns.

\subsection{Beyond the nPVI}
Generality is one of the main benefits of the anisochrony measure introduced above: it is defined for arbitrarily long segments. 
The nPVI only quantifies first order isochrony: when \emph{two} successive intervals have the same length. 
Figure~\ref{fig:cluster-transition-network} is an example of a dataset with substantial length-2 isochrony, as can be seen from the ratio plot at the top. 
Still, segments of \emph{three} identical intervals are completely absent, as can be seen in Figure~\ref{fig:3d-pat-dur}.
This is a distinction length-3 anisochrony can make, while the nPVI cannot. 
Put differently, the anisochrony gives a stricter assessment of isochrony than the nPVI when longer segment lengths are used.

The measure can be extended in a second direction: by changing the ‘reference’ pattern. 
Anisochrony measures the average normalized distance to one particular reference pattern, namely the isochronous one. 
Depending on the research question, other references may be more relevant. 
For example, perhaps the average distance to the $7 : 2 : 3$ \emph{maraka} rhythm could be an informative measure when studying music from Mali \citep[cf.][]{Jacoby2024NatHumBehav}.
Such measures will remain crude summaries and should probably be complemented by at least a visualization of the pattern distribution. 
A similar point is made by \citet{Jadoul2025Vibration}, who go on to suggest altogether different measures, such as the entropy of the pattern distribution—a suggestion left for future work. 

\section{Discussion}

In this paper, I proposed analyzing rhythmic data by studying its segments. 
This simple conceptual framework turned out to be remarkably productive. 
Decomposing segments into patterns and durations maps them to a convenient mathematical space, the rhythm simplex, that lends itself well to visualization. 
In particular, I introduced pattern-duration plots that effectively summarized regularities in instrument onsets, especially when overlaid with a cluster transition network.
Compared to existing alternatives, pattern-duration plots make efficient use of visual space, their axes represent meaningful quantities, and their marginal density plots are directly informative.
Finally, a number of quantitative measures are naturally explained by this framework. 
Rhythm ratios are simply length-2 patterns, and the nPVI is an average distance from isochrony, perhaps better replaced with the more general and normalized anisochrony measure. 

Still, the main contribution is neither the plotting method nor the measures, but the simple concepts of segments and patterns: they tie together the work on which this paper builds and they may simplify our reasoning about rhythmic data. 
More broadly, this paper argues for visualizing rhythmic data before resorting to quantitative measures such as the nPVI,
and underlines that the nPVI or its proposed replacement, the anisochrony, should only be used when one actually needs to assess isochrony.
Otherwise, the framework provides the conceptual tools for developing more appropriate measures tailored to a particular study.
Future work could explore extending the framework, for example by combining segments of different lengths, or studying segments that skip one or more intervals (i.e.~skip-grams). 

Since the anisochrony effectively generalizes the nPVI, the criticism of the nPVI raised by \citet{Condit-Schultz2019MusicPercept} may carry over to the present paper. 
However, a central part of the critique concerns the application of an average to discrete intervals in notated rhythm. 
For the continuous data analyzed here, this seems less problematic. 
More generally, \citet{Condit-Schultz2019MusicPercept} argues for using standard measures like the nPVI critically, advocates a plurality of measures, and stresses the importance of musically relevant rhythmic features over general measures. 
All of this, in fact, aligns well with the kind of rhythmic segment analyses I argue for here.

Throughout this paper, I have avoided cognitive claims and assumptions, so that the framework remains applicable to any kind of rhythmic data. 
In particular, the pattern distance is not intended as a model of perceived dissimilarity:
to the extent that rhythm is perceived categorically, perceived dissimilarity might increase across category boundaries \citep{Desain2003Perception}, while the pattern distance remains unaffected.
This may actually be desirable, since operationalizing such boundary effects also requires a neutral distance measure.
An aspect that may however lend itself more directly to cognitive interpretation is the distinction between patterns and durations (tempo).

This paper also introduced the concept of quantality to describe rhythmic data that approximately falls on a regular grid. 
Throughout the paper, the quantum was taken to be constant, which implies a constant tempo. 
This is an unrealistic assumption for a lot of music, and so future work could consider expressing segment durations in terms of a time-varying quantum. 
After all, the concept itself seems productive. 
Pattern-duration plots, for example, are more insightful when all integer-ratio segments are annotated, which requires a quantum. 
Indeed, quantality turns out to be very closely connected to the presence of small-integer-ratio rhythms—a topic of much recent \emph{perceptual} research \citep{Jacoby2024NatHumBehav,Jacoby2017CurrBiol,Ravignani2018FrontComputNeurosci}.
The two notions appear to be two sides of the same coin. 
This suggests that questions about integer ratios can be reformulated as questions about quantality—connecting the present work to fundamental questions about the structure of musical rhythm.

\section*{Data and code availability}
\addcontentsline{toc}{section}{Data and code availability}

All data and code used in this paper have been deposited at \url{https://doi.org/10.5281/zenodo.17753529}.

\addcontentsline{toc}{section}{References}

\end{document}